\documentclass[aip,super]{revtex4-2}
\usepackage{graphicx,amsmath}
\usepackage{amsmath,amssymb}
\usepackage[matrix,frame,arrow]{xy}
\usepackage{color}
\usepackage[vcentermath]{youngtab}
\usepackage{braket}
\usepackage{appendix}
\usepackage{siunitx}
\usepackage{mathtools}
\usepackage{url}
\usepackage{dsfont}
\usepackage{verbatim}
\usepackage{rotating}
\usepackage[ruled,vlined]{algorithm2e}
\usepackage{qcircuit}
\usepackage{subfig}

\makeatletter
\newcommand{\mathleft}{\@fleqntrue\@mathmargin0pt}
\newcommand{\mathcenter}{\@fleqnfalse}
\makeatother

\begin{document}
\date{\today}
\title{Domain-Specific Compilers for Dynamic Simulations of Quantum Materials on Quantum Computers}
\author{Lindsay Bassman}
\affiliation{Collaboratory for Advanced Computing and Simulations, University of Southern California, Los Angeles, CA 90089, United States of America}
\author{Sahil Gulania}
\affiliation{Department of Chemistry, University of Southern California, Los Angeles, CA 90089, United States of America}
\author{Connor Powers}
\affiliation{Collaboratory for Advanced Computing and Simulations, University of Southern California, Los Angeles, CA 90089, United States of America}
\author{Rongpeng Li}
\affiliation{Department of Physics, University of Southern California, Los Angeles, CA 90089, United States of America}
\author{Thomas Linker}
\affiliation{Collaboratory for Advanced Computing and Simulations, University of Southern California, Los Angeles, CA 90089, United States of America}
\author{Kuang Liu}
\affiliation{Collaboratory for Advanced Computing and Simulations, University of Southern California, Los Angeles, CA 90089, United States of America}
\author{T. K. Satish Kumar}
\affiliation{Department of Computer Science, University of Southern California, Los Angeles, CA 90089, United States of America}
\author{Rajiv K. Kalia}
\affiliation{Collaboratory for Advanced Computing and Simulations, University of Southern California, Los Angeles, CA 90089, United States of America}
\author{Aiichiro Nakano}
\affiliation{Collaboratory for Advanced Computing and Simulations, University of Southern California, Los Angeles, CA 90089, United States of America}
\author{Priya Vashishta}
\affiliation{Collaboratory for Advanced Computing and Simulations, University of Southern California, Los Angeles, CA 90089, United States of America}

\begin{abstract}
Simulation of the dynamics of quantum materials is emerging as a promising scientific application for noisy intermediate-scale quantum (NISQ) computers.  Due to their high gate-error rates and short decoherence times, however, NISQ computers can only produce high-fidelity results for those quantum circuits smaller than some given circuit size.  Dynamic simulations, therefore, pose a challenge as current algorithms produce circuits that grow in size with each subsequent time-step of the simulation.  This underscores the crucial role of quantum circuit compilers to produce executable quantum circuits of minimal size, thereby maximizing the range of physical phenomena that can be studied within the NISQ fidelity budget.  Here, we present two domain-specific quantum circuit compilers for the Rigetti and IBM quantum computers, specifically designed to compile circuits simulating dynamics under a special class of time-dependent Hamiltonians.  The compilers outperform state-of-the-art general-purpose compilers in terms of circuit size reduction by around 25-30\% as well as wall-clock compilation time by around 40\% (dependent on system size and simulation time-step).  Drawing on heuristic techniques commonly used in artificial intelligence, both compilers scale well with simulation time-step and system size.  Code for both compilers is included to enhance the results of dynamic simulations for future researchers.  We anticipate that our domain-specific compilers will enable dynamic simulations of quantum materials on near-future NISQ computers that would not otherwise be possible with general-purpose compilers.  
\end{abstract}
\maketitle

\section{Introduction}
Quantum computers promise to solve certain classes of problems that are intractable on the most advanced classical computers by massively reducing either time-to-solution, computational resource requirements, or both.  By storing and processing information on quantum bits, or qubits, quantum computers take advantage of quantum mechanical phenomena, such as superposition and entanglement, to beat performance of their classical counterparts.  Nearly forty years after their theoretical conception as universal quantum simulators\cite{feyn82}, quantum computers are beginning to produce early successes in simulating quantum systems.  However, currently available and near-future quantum computers, commonly known as noisy intermediate-scale quantum (NISQ) computers, suffer from high gate- and measurement-error rates, as well as qubit decoherence\cite{presk18}.   Furthermore, due to their small numbers of qubits, NISQ computers are unable to exploit robust error-correcting schemes which are expected to give rise to fault-tolerant quantum computers in the future.  As a result, the only physical systems that have been simulated on quantum computers to date\cite{lany10,kand17,lierta18,colles18,hempel18,lamm18} are too small to see a quantum advantage (i.e., these systems can still be simulated on classical computers).  While such simulations have provided encouraging proofs-of-concept, the next great challenge is to learn new physics by simulating a quantum material on a quantum computer, which cannot feasibly be simulated on any classical computer.

A particularly promising route to discovering new physics with NISQ computers is the dynamic simulation of strongly-correlated quantum systems.  On the one hand, the complexity of such simulations on classical computers grows exponentially with the number of particles in the system, quickly making simulations of even modestly-sized systems impossible to run on state-of-the-art classical supercomputers.  On the other hand, only modestly-sized systems are required to study the dynamics of various physical models, making it possible to fit such simulations onto NISQ computers.   Together, this means there may exist a “goldilocks” quantum system, too large for classical computers but small enough for NISQ computers to simulate, with which to achieve physical quantum supremacy.

Dynamic simulations are carried out on quantum computers by creating a different quantum circuit, or sequence of quantum logic gates carried out on each qubit, for each time-step of the simulation.  Current algorithms generally produce quantum circuits that grow in size with each time-step~\cite{wiebe11,martinez2016real,smith2019simulating}.  Since NISQ computers can only execute circuits up to a certain size with high fidelity, due to high gate-error rates and low qubit decoherence times, this puts a limit on the number of time-steps a NISQ computer can feasibly simulate.  It is therefore crucial to develop quantum circuit compilers that can minimize circuit sizes enough to enable dynamic simulations on NISQ computers.  

Still in their infancy, quantum circuit compilers are an integral part of the quantum computing software stack \cite{chong2017programming}.  Their function is to transform a high-level quantum circuit, usually the output of an algorithm, into a circuit that can be executed on a quantum computer.  High-level quantum circuits are defined by a set of arbitrary quantum logic gates, each of which can be represented by a unitary matrix, acting on different subsets of qubits in the system.  Quantum computers, however, are only designed with the ability to perform a small, universal set of one- and two-qubit gates, called their native gate set.  Products of these native gates can be designed to be equivalent to any N-qubit gate, and thus can be used to execute any high-level quantum circuit.  A complicating factor in the NISQ-era is that different quantum computers have different native gate sets.  Therefore, the executable quantum circuit produced by the quantum circuit compiler must only consist of those native gates specific to the quantum machine upon which it will be executed. 

On top of mapping high-level circuits to native-gate circuits, the quantum circuit compiler is further expected to optimize the native-gate circuit, which in the NISQ-era equates to circuit size minimization.  There are two main methods for quantum circuit optimization.  The first method, which we call the minimal universal circuit (MUC) method, requires the construction of an MUC for each integer number of qubits.  The circuit is universal in the sense that it can implement arbitrary unitary evolution of the qubits, and is minimal in the sense that it uses the fewest number of gates to implement the unitary evolution.  Each MUC has an associated set of parameters whose values must be solved for based on the action the corresponding high-level circuit is meant to execute.  The method works by representing all gates of the high-level circuit in their matrix representation and multiplying them into one $2^N \times 2^N$ unitary matrix, where $N$ is the total number of qubits in the circuit.  This matrix is then used to determine the values of the parameters in the MUC.  Note that while the values of the parameters in the MUC change from time-step to time-step, the size of the MUC remains the same, irrespective of how many gates were present in the original high-level circuit.  If the MUC circuit for $N$ qubits is small enough for high-fidelity results on the NISQ computer, then one can, in principle, simulate evolution of a system of $N$ qubits out to an arbitrary number of time-steps.  

Despite the advantage of constant-size circuits, the MUC method has two substantial drawbacks.  First, the MUC itself must be created for each qubit number.  Thus far, this has only been studied for two and three qubits\cite{shende2004minimal, bullock2003arbitrary, vidal2004universal, vatan2004optimal, vatan2004realization}.  Second, for an $N$-qubit circuit, the compiler must compute a 2$^{N}\times$2$^{N}$ unitary matrix, causing the memory resources of this method to grow exponentially with system size.  Together, these issues render the MUC method unsuitable for compilation of the larger-scale simulation circuits required for new discoveries on NISQ computers.

The second method of compilation, which we call the transform and reduce (TR) method, involves using algebraic identities to substitute subsets of gates in the circuit with alternate (e.g., smaller, native) sets of gates.  The method works by scanning through a circuit, searching for sets of gates onto which to apply the algebraic identities.  While this method decreases circuit size for each time-step, the executable circuits still grow with increasing time-step.  The advantage is that this method can be made to scale efficiently with system size, giving it the potential to work with circuits simulating large enough systems to demonstrate physical quantum supremacy.  However, the large number of identities, along with the enormous number of permutations in which they can be applied, makes optimizing this method an NP-hard~\cite{herr17, botea2018} problem.  

Since as few as $\mathcal{O}(1)$ gates can make a significant difference in the fidelity of a quantum circuit executed on a NISQ computer, the importance of quantum circuit optimization is paramount.  Successfully discovering new physics with NISQ computers will require larger numbers of qubits, making the MUC method untenable due to its poor scaling behavior with system size.  Therefore, near-future NISQ simulations require the scalability of the TR method.  To address the difficulty of optimizing the TR method, we propose developing compilers for a specific native gate set and a specific class of quantum circuits.  Creating such {\em domain-specific} compilers allows the designer to take advantage of the structure of the specific problem to develop heuristics that enable better and faster optimization compared to a general-purpose compiler.

Here, we present two domain-specific (DS) quantum circuit compilers for the cloud-accessible NISQ computers provided by Rigetti and IBM.  Each DS compiler compiles circuits into the native gate set of one of these machines.  Both are specifically designed for compiling circuits that simulate time evolution under a special class of Hamiltonians known as the transverse field Ising model (TFIM) with a time-dependent transverse field \cite{lo1990ising, chakrabarti2008quantum}.  The TFIM is considered a quintessential model for studying quantum phase transitions\cite{suzuki2012quantum}, as well as myriad condensed matter systems, such as ferroelectrics~\cite{blinc1979ising} and magnetic spin glasses~\cite{wu1991classical}.  When the transverse field is time-dependent, non-equilibrium effects such as dynamic phase transitions and quantum hysteresis can be studied \cite{tome1990dynamic,acharyya1995response,acharyya1998nonequilibrium,sides1998kinetic}.  Simulation of a time-dependent TFIM spin system on a NISQ computer is therefore an excellent candidate for the discovery of new physics and we choose this important model to demonstrate how NISQ-era dynamic simulations can benefit from DS compilers.  The DS compilers are able to reduce quantum circuit sizes by around 25-30\% and compilation times by around 40\% compared to state-of-the-art general-purpose compilers (values vary depending on system size and simulation time-step).  We have made the code for both DS compilers available to extend the range of feasible dynamic simulations for future researchers on NISQ computers \cite{githubDSQC}.  We anticipate that the DS compilers will allow for successful simulations of larger material systems on near-future NISQ computers that could not otherwise be achieved with general-purpose compilers, due to the shorter circuit sizes they produce.

\section{Results}
\subsection{Importance of Optimized Quantum Circuit Compilers}
\begin{figure}[h!]
    \centering
    \includegraphics[scale=0.9]{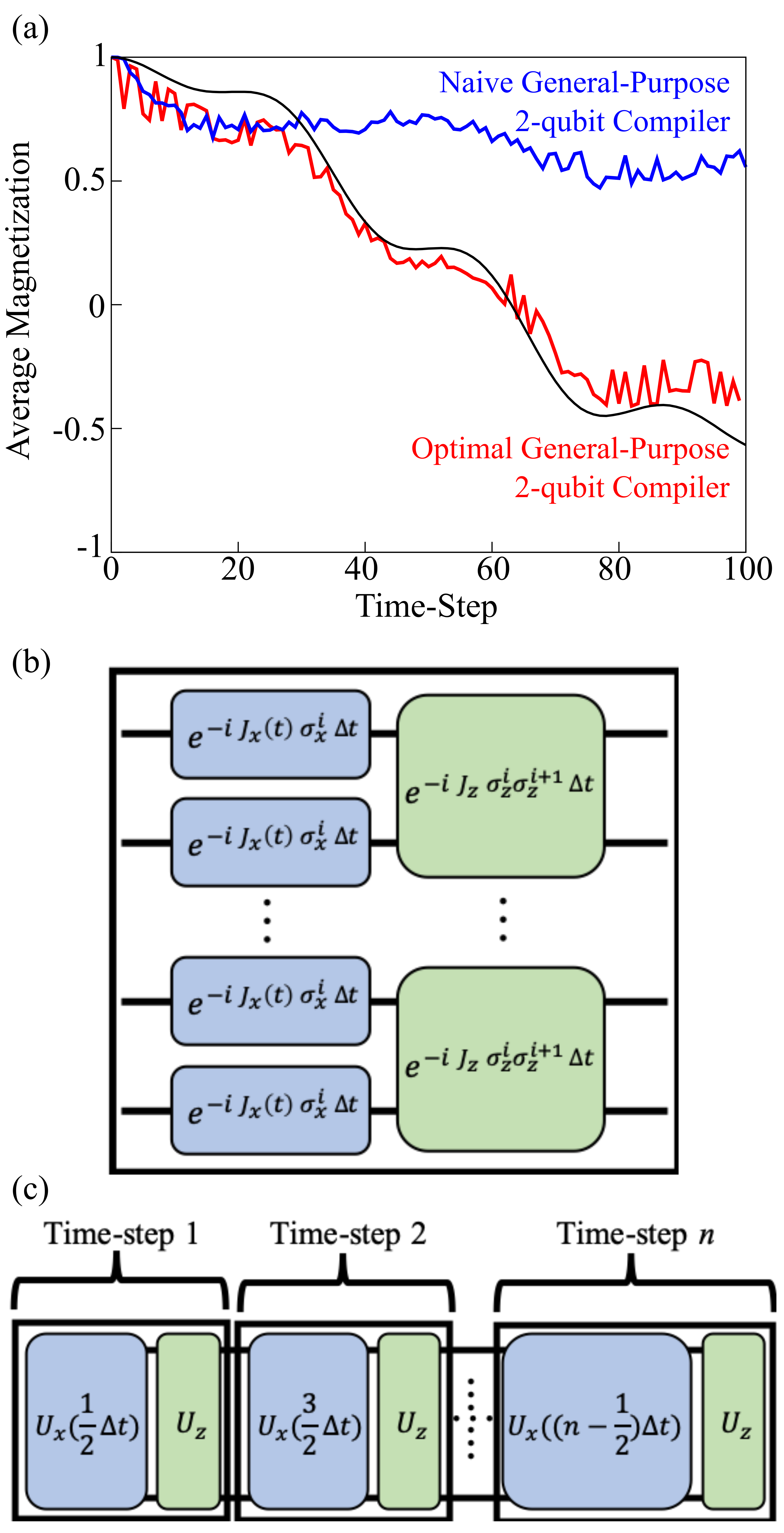}
    \caption{The importance of quantum circuit compilation and the general structure of TFIM circuits.  (a) Comparison of simulation results from Rigetti's quantum computer for a 2-qubit system compiled naively (blue) versus using the optimized MUC method (red).  (b) Block of gates representing the evolution of the system by one time-step.  (c) High-level quantum circuit depicting the evolution of the system by $n$ time-steps. }
    \label{fig:trotter}
\end{figure}

To showcase the importance of optimized quantum circuit compilers, we begin by showing results from simulations performed on Rigetti's quantum computer for a two-qubit system using a general-purpose compiler with and without optimization. For two-qubit systems, Rigetti’s and IBM's optimized general-purpose compilers use the MUC method of compilation, allowing for constant-depth circuits for all time-steps.  Without optimization, the general-purpose compilers use a naïve method that simply translates the high-level circuits directly into native gate circuits, thereby creating executable circuits that grow with each time-step.  The stark difference in results from simulations using the naïve general-purpose two-qubit compiler (blue) versus the optimized general-purpose two-qubit compiler (red) is shown in Figure \ref{fig:trotter}a.  
Since the optimized, constant-depth circuits are small enough to produce high-fidelity results, the NISQ computer is able to simulate the system out to arbitrary time-steps, as can be seen by the close correspondence of the quantum computer results (red) to the numerical results (black) from a simulated quantum computer.  The naïve compiler initially produces circuits that are small enough to achieve close correspondence with the numerical results.  After a certain number of time-steps, however, it produces circuits that are too large, generating too much error and leading to results that drastically diverge from the ground truth.  As all factors are held constant between the two simulations except the compiler, these results emphasize the importance of quantum circuit compilers for achieving high-fidelity results on NISQ computers.  

\subsection{Domain-Specific Compilers for TFIM Circuits}
In this article, we present two DS compilers for compilation of circuits simulating spin-evolution under a time-dependent TFIM Hamiltonian into the native gate set of either Rigetti's or IBM's quantum computer.  The TFIM Hamiltonian, defined by 
\begin{equation} \label{hamiltonian}
    H(t) = -J_{z}\sum_{i=1}^{N-1} \sigma_{i}^{z}\sigma_{i+1}^{z} - J_{x}(t) \sum_{i=1}^{N} \sigma_{i}^{x}
\end{equation}
models a system of spins with nearest-neighbor exchange interactions of strength $J_z$, in the presence of a transverse magnetic field with a time-dependent amplitude defined by $J_{x}(t)$.  Here, $\sigma_{i}^{\alpha}$ is the $\alpha$-Pauli matrix acting on qubit $i$.  The system can be simulated a quantum computer by mapping the spins of the TFIM to the spins of the qubits.  The dynamics of the system is simulated by applying to the qubits a sequence of quantum logic gates that is equivalent to time-evolution under Hamiltonian (1), given by the time-ordered unitary operator $U(t)\equiv U(0,t)=\mathcal{T} exp(-i\int_{0}^{t}H(t)dt)$
in the atomic unit.  Deriving this sequence of quantum logic gates involves discretizing time and applying the Trotter approximation \cite{trotter1959product} to $U(t)$ to arrive at the following approximate time-evolution operator: 
\begin{equation}\label{unitary}
    U(n\Delta t) = \prod_{j=0}^{n-1} e^{-iH_{x}((j+1/2)\Delta t)\Delta t} e^{iH_{z}\Delta t} + \mathcal{O}(\Delta t)
\end{equation}
where, $H_x (t)=-J_{X}(t)\sum_{i=1}^{N} \sigma_{i}^{x}$ and $H_z = -J_z \sum_{i=1}^{N-1} \sigma_{i}^{z} \sigma_{i+1}^{z}$.  More details can be found in the Methods section. 
 
Dynamic simulations are performed by creating a different quantum circuit for each time-step of the total simulation, where the circuit for time-step $n$ simulates evolution from time $t=0$ to $t=n\Delta t$ by implementing $U(n\Delta t)$.  Examination of Equation \ref{unitary} shows that the circuit for time-step $n$ consists of the product of a number of operators proportional to $n$.  Circuits, therefore, grow in size with increasing time-step.  The general form of the circuits is shown in Figures \ref{fig:trotter}b and c.  

Figure \ref{fig:trotter}b shows a block of gates enacting evolution of the system by one time-step $\Delta t$.  The blue boxes represent single-qubit operators that apply the transverse magnetic field to each qubit for a time $\Delta t$, while the green boxes represent two-qubit operators that apply the exchange interaction between nearest-neighbor qubits for a time $\Delta t$.  Figure \ref{fig:trotter}c shows a high-level circuit diagram for simulating evolution of the system for a total time $n\Delta t$.  Note how the block of gates shown in Figure \ref{fig:trotter}b is repeated $n$ times in Figure \ref{fig:trotter}c, although the parameters of each block vary per time-step due to the time-dependence of the Hamiltonian.  Nonetheless, a regular structure appears in the circuits for TFIM simulations, which can be exploited for developing heuristics for a DS quantum circuit compiler.

As the MUC compilation method cannot scale to systems large enough to demonstrate a quantum advantage, we use the TR method for development of the DS compilers.  The DS compilers are designed to take as input high-level circuits containing gates from the set $\{H, RZ(\theta), CNOT\}$, based on the algorithm given in ref\cite{bassman2020towards} for TFIM simulation.  The target native gate sets are $\{RX(\pm \frac{\pi}{2}), RZ(\theta), CZ\}$ for Rigetti's quantum computer, and $\{RX(\frac{\pi}{2}), RZ(\theta), CNOT\}$ for IBM's quantum computer, for arbitrary angle $\theta$.  In all gate sets, $H$ is the Hadamard gate; $RX(\theta)$ and $RZ(\theta)$ are single-qubit rotation gates that rotate the qubit about the $x$-axis and $z$-axis, respectively, by an angle $\theta$; $CNOT$ is a two-qubit controlled-not gate; and $CZ$ is a two-qubit controlled-$Z$ gate\cite{nielsen2002quantum}.

All algebraic identities used for gate transformation and reduction in the DS compilers are detailed in Table \ref{gate_identites}.  The left-hand column numbers the identities for identification.  The middle column gives sets of gates commonly found in uncompiled TFIM circuits, while the right-hand column shows the corresponding equivalent to each set of gates.  Identities 1 and 2 are used for converting high-level gates into native gates.  Note that while they do not reduce the gate count, they are essential for producing circuits that are executable on the quantum processors.  Identities 3-5 are gate-altering and re-organizing identities, which prime circuits for further reductions.  Note that identity 5, in particular, leverages knowledge of TFIM circuits by identifying a complex set of gates on which to apply a re-organizing identity. Finally, identities 6-10 are simplifying identities that are used purely to reduce gate count.  Note that if $\theta_i = -\theta_j$ in identities 7-10, then a rotation by zero radians results, which can be completely removed from the circuit.
\begin{table}[h!]
\caption{Algebraic identities for gate sets commonly found in TFIM circuits.}
\begin{tabular}{|c|c|c|}
\hline 
No. & Common Gate Set in TFIM Circuits & Equivalent   \\ \hline
1 & $\Qcircuit @C=1em @R=.7em {
    & \ctrl{1} & \qw\\ 
    &\targ     &  \qw
    }$ 
    & $\Qcircuit @C=1em @R=.7em {
    & \qw &  \gate{Z} & \qw & \qw\\ 
    & \gate{H}  &\ctrl{-1}  &  \gate{H} &  \qw }$ \\ \hline
2 & $\Qcircuit @C=1em @R=.7em {
    & \gate{H}& \qw
    }$ & 
    $\Qcircuit @C=1em @R=.7em {
    & \gate{RZ(\frac{\pi}{2})}  & \gate{RX(\frac{\pi}{2})} & \gate{RZ(\frac{\pi}{2})} & \qw } $ \\ \hline
3 & $\Qcircuit @C=1em @R=.7em {
    & \ctrl{1} & \qw\\ 
    &\targ     &  \qw
    }$ 
    & $\Qcircuit @C=1em @R=.7em {
    & \gate{H} &  \targ & \gate{H} & \qw\\ 
    & \gate{H} & \ctrl{-1}  &  \gate{H} & \qw }$ \\ \hline
4 & $    \Qcircuit @C=1em @R=.7em {
    & \gate{RX(\theta_{i})} & \gate{RZ(\pi)} & \qw
    } $  & 
    $\Qcircuit @C=1em @R=.7em {
    & \gate{RZ(-\pi)} & \gate{RX(-\theta_{i})}  &\qw} $\\ \hline
5 & {\footnotesize$    \Qcircuit @C=0.3em @R=.2em {
    & \gate{Z}  & \qw & \qw & \qw &\gate{Z}  & \gate{RX(-\frac{\pi}{2})} & \qw  \\
    &\ctrl{-1}    & \gate{RX(\frac{\pi}{2})} & \gate{RZ(\theta_{i})} & \gate{RX(-\frac{\pi}{2})} & \ctrl{-1}
    & \gate{RX(\frac{\pi}{2})} & \qw
    }$} &
    {\footnotesize$
    \Qcircuit @C=0.3em @R=.2em {
    & \gate{RX(-\frac{\pi}{2})}&\gate{Z}      & \gate{RX(\frac{\pi}{2})} & \gate{RZ(\theta_{i})} & \gate{RX(-\frac{\pi}{2})} &\gate{Z}  & \qw \\
    & \gate{RX(\frac{\pi}{2})} & \ctrl{-1} & \qw & \qw & \qw  
    & \ctrl{-1} & \qw}$}\\ \hline
6 & $\Qcircuit @C=1em @R=.7em {
    & \gate{H}& \gate{H} & \qw
    }$ &  
    $\Qcircuit @C=1em @R=.7em {
    & \qw & \qw} $ \\ \hline
7 & $    \Qcircuit @C=1em @R=.7em {
    & \gate{RX(\theta_{i})} & \gate{RX(\theta_{j})} & \qw
    } $&
    $
    \Qcircuit @C=1em @R=.7em {
    & \gate{RX(\theta_{i}+\theta_{j})} & \qw }$\\ \hline
8 & $    \Qcircuit @C=1em @R=.7em {
    & \gate{RZ(\theta_{i})} & \gate{RZ(\theta_{j})} & \qw
    } $&
    $
    \Qcircuit @C=1em @R=.7em {
    & \gate{RZ(\theta_{i}+\theta_{j})} & \qw }$\\ \hline
9 & $ \Qcircuit @C=1em @R=.7em {
    & \gate{RZ(\theta_{i})} & \ctrl{1} 
    & \gate{RZ(\theta_{j})} & \qw\\
    &\qw & \gate{Z} &  \qw & \qw
    } $
    & $ \Qcircuit @C=1em @R=.7em {
    &\ctrl{1} & \gate{RZ(\theta_{i}+\theta_{j})} & \qw\\
    & \gate{Z} &  \qw & \qw } $\\ \hline
10 & $ \Qcircuit @C=1em @R=.7em {
    & \gate{RZ(\theta_{i})} & \ctrl{1}
    & \gate{RZ(\theta_{j})} & \qw\\
    &\qw & \targ &  \qw & \qw
    } $
    & $ \Qcircuit @C=1em @R=.7em {
    & \ctrl{1} & \gate{RZ(\theta_{i}+\theta_{j})} & \qw\\
    &\targ &  \qw & \qw } $\\ \hline
\end{tabular}
\label{gate_identites}
\end{table}
Each compiler uses a different subset of these identities, chosen for targeting different native gate sets.  Not only is it important to implement identities specifically useful for a given circuit type and native gate set, it is also important to use knowledge of the circuit structure to develop heuristics for optimal ordering of identity application.  Different permutations of identity application can lead to varied resultant circuits.  Details on how each DS compiler is implemented, including pseudocode, are described in the subsections below.
 
\subsection{Rigetti Native Gate Set Domain-Specific Compiler}
Algorithm \ref{rigetti_pseudocode} shows pseudocode detailing the order in which a subset of the identities from Table \ref{gate_identites} are applied in the DS compiler for Rigetti's native gate set.  The first loop uses identity 1 to transform all $CNOT$ gates into a set of Hadamard and $CZ$ gates, since $CNOT$ gates are not in Rigetti’s native gate set.  The second loop uses identity 6 to remove consecutive pairs of Hadamard gates acting on the same qubit introduced by the previous loop.  The third loop uses identities 2, 8, and 9 to convert Hadarmard gates to native gates and compress RZ gates on the same qubit that are either adjacent or on opposite sides the control bit of a $CZ$ gate.  By searching for a common motif in TFIM circuits, the third loop is able to carry out multiple identities at once, speeding up the wall-clock compile time.  The fourth loop uses identity 2 to transform all remaining Hadamard gates into native gates, as Hadamard gates are not in Rigetti’s native gate set.  At this point, the circuit is comprised solely of native gates and in principle can be executed on the quantum processor.  However, since large circuits have poor fidelity on NISQ computers, the remainder of the compilation process acts to minimize the number of gates in the circuit.  The fifth loop uses identity 8 to concatenate neighboring $RZ$ gates on the same qubit.  The sixth and seventh loops work in concert to apply identities 5 and 7 for further gate reductions.  The eighth and ninth loops use identities 8 and 4, respectively, to reduce the number of $RZ$ gates.  Finally, not shown in the pseudocode, is the removal of any trailing gates at the end of the circuit that only alter the phase of the system, and thus do not affect any measurement gates that immediately follow. 

Of particular interest in Algorithm 1 is the relationship between the sixth loop, which performs gate rearrangement, and its inner loop (seventh loop), which performs gate reduction. The inner loop concatenates pairs of adjacent rotation gates acting on the same qubit along the same axis, reducing two gates to one. The outer loop alters gate placement but does not change the number of gates. We observe that before each subsequent iteration of the outer loop, the inner loop is applied until no more reductions are possible. In this way, gate reductions enabled by the inner loop are carried out fully before each subsequent step of gate rearrangement. This strategy is similar to the ones used in heuristic search under the name of {\em unit propagation}. In heuristic search, and particularly in solvers built for Satisfiability (SAT) problems, unit propagation is very useful~\cite{la97}. It reduces two disjuncts in the search space to a single disjunct whenever such a reduction is valid (analogous to reducing two consecutive RZ gates to one gate). Calling it in each step of the search serves as an efficient and effective lookahead technique~\cite{edelkamp2011heuristic}.
\begin{algorithm}[!h]
\caption{Pseudocode for domain-specific compilation of TFIM circuits into Rigetti's native gate set.}
 Read in high-level gates\;
 {\scriptsize
 
 1: \While{$ \Qcircuit @C=1em @R=.7em {
    & \ctrl{1} & \qw\\ 
    &\targ     &  \qw
    }$ exists } {\vspace{0.5cm}
  replace it with $\Qcircuit @C=1em @R=.7em {
    & \qw &  \gate{Z} & \qw & \qw\\ 
    & \gate{H}  &\ctrl{-1}  &  \gate{H} &  \qw }$\
}
 2: \While{$\Qcircuit @C=1em @R=.7em {
    & \gate{H}& \gate{H} & \qw
    }$ exists }{\vspace{0.5cm}
    replace it with
  $\Qcircuit @C=1em @R=.7em {
    & \qw & \qw & \qw} $
}

 3: \While{{$    \Qcircuit @C=0.5em @R=.7em {
    & \gate{Z}  & \qw & \qw & \qw &\gate{Z}  & \qw  \\
    &\ctrl{-1}     & \gate{H} & \gate{RZ(\theta_{i})} & \gate{H} & \ctrl{-1}
    & \qw
    }$} exists }{\vspace{0.5cm}
    replace it with
  $    \Qcircuit @C=0.5em @R=.7em {
    & \qw & \gate{Z}  & \qw & \qw & \qw &\gate{Z}  & \qw & \qw \\
    &\gate{RZ(\frac{\pi}{2})} &\ctrl{-1}     & \gate{RX(\frac{\pi}{2})} & \gate{RZ(\theta_{i})} & \gate{RX(-\frac{\pi}{2})} & \ctrl{-1}
    &\gate{RZ(\frac{-\pi}{2})} &\qw
    }$
}

 4: \While{$\Qcircuit @C=1em @R=.7em {
    & \gate{H}& \qw
    }$ exists }{\vspace{0.5cm}
    replace it with
  $\Qcircuit @C=1em @R=.7em {
    & \gate{RZ(\frac{\pi}{2})}  & \gate{RX(\frac{\pi}{2})} & \gate{RZ(\frac{\pi}{2})} & \qw } $ 
}

 5: \While{$    \Qcircuit @C=1em @R=.7em {
    & \gate{RZ(\theta_{i})} & \gate{RZ(\theta_{j})} & \qw
    } $ exists  }{\vspace{0.5cm}
    replace it with
    $
    \Qcircuit @C=1em @R=.7em {
    & \gate{RZ(\theta_{i}+\theta_{j})} & \qw }$
    }

 6: \While{{$    \Qcircuit @C=0.5em @R=.7em {
    & \gate{Z}  & \qw & \qw & \qw &\gate{Z}  & \gate{RX(-\frac{\pi}{2})} & \qw  \\
    &\ctrl{-1}     & \gate{RX(\frac{\pi}{2})} & \gate{RZ(\theta_{i})} & \gate{RX(-\frac{\pi}{2})} & \ctrl{-1}
    & \gate{RX(\frac{\pi}{2})} & \qw
    }$ exists }}
    {\vspace{0.5cm} replace it with {$
    \Qcircuit @C=0.5em @R=.7em {
    & \gate{RX(-\frac{\pi}{2})}&\gate{Z}      & \gate{RX(\frac{\pi}{2})} & \gate{RZ(\theta_{i})} & \gate{RX(-\frac{\pi}{2})} &\gate{Z}  & \qw \\
    & \gate{RX(\frac{\pi}{2})} & \ctrl{-1} & \qw & \qw & \qw  
    & \ctrl{-1} & \qw}$}
    
 7: \While{$    \Qcircuit @C=1em @R=.7em {
    & \gate{RX(\theta_{i})} & \gate{RX(\theta_{j})} & \qw
    } $ exists }{\vspace{0.5cm}
    replace it with
    $
    \Qcircuit @C=1em @R=.7em {
    & \gate{RX(\theta_{i}+\theta_{j})} & \qw }$
     }
    }

 8: \While{$    \Qcircuit @C=1em @R=.7em {
    & \gate{RZ(\theta_{i})} & \gate{RZ(\theta_{j})} & \qw
    } $ exists  }{\vspace{0.5cm}
    replace it with
    $
    \Qcircuit @C=1em @R=.7em {
    & \gate{RZ(\theta_{i}+\theta_{j})} & \qw }$
    }

 9: \While{$    \Qcircuit @C=1em @R=.7em {
    & \gate{RZ(\theta)} &\gate{RX(\frac{\pi}{2})} & \gate{RZ(\pi)} & \qw
    } $ exists }{\vspace{0.5cm}
    replace it with
  $\Qcircuit @C=1em @R=.7em {
    & \gate{RZ(\theta - \pi)} & \gate{RX(-\frac{\pi}{2})}  &\qw} $  
} 
}
\label{rigetti_pseudocode}
\end{algorithm}

\begin{figure}[h!]
    \centering
    \includegraphics[scale=0.85]{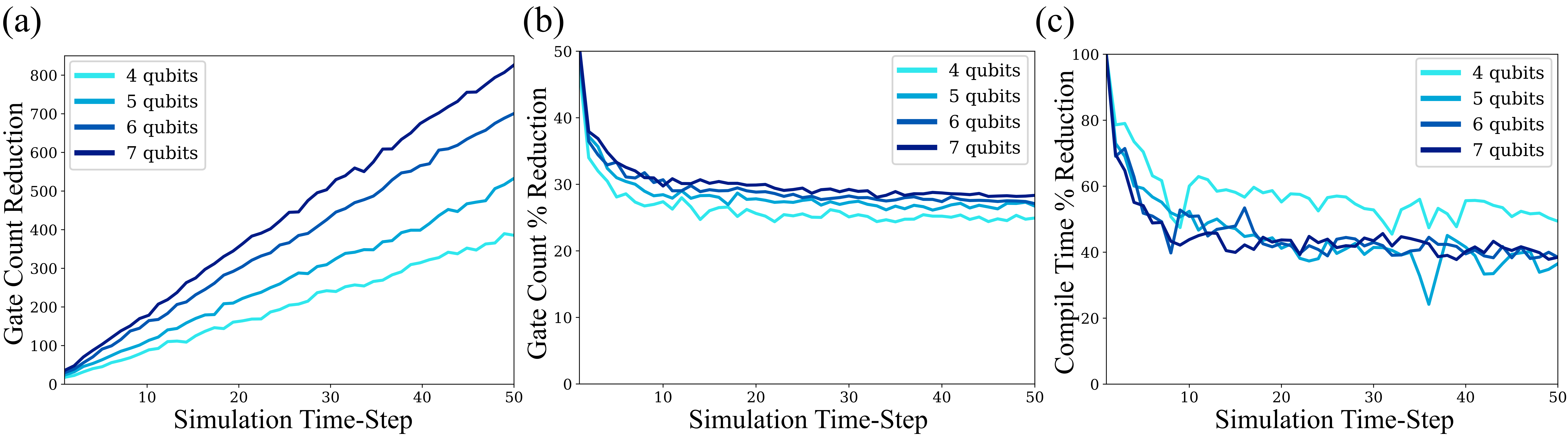}
    \caption{Performance comparison of the DS compiler to Rigetti's general-purpose compiler. (a) Absolute reduction in gate count using the DS compiler over Rigetti's general-purpose compiler, for varying system size.  (b) Percent reduction in gate count using the DS compiler with respect to Rigetti-compiled circuit size, for varying system size.  (c) Percent reduction in wall-clock compilation time using the DS compiler with respect to Rigetti-compiled circuit size, for varying system size.}
    \label{fig:rigetti_performance}
\end{figure}

To test the performance of this compiler, we constructed an algorithm to create a sequence of high-level circuits that simulate evolution through time of a system of spins under a TFIM Hamiltonian, as given in Equation \ref{hamiltonian}.  Details of the algorithm can be found in the Methods section.  The high-level circuits were fed into the DS compiler as well as Rigetti's general-purpose compiler, and we compared performance based on the number of native gates in the compiled circuits and the wall-clock time for compilation for each time-step, for systems with varying numbers of qubits.  Results can be found in Figure \ref{fig:rigetti_performance}.  
Figure \ref{fig:rigetti_performance}a shows the absolute native gate count reduction when using the DS compiler over Rigetti's general-purpose compiler.  As shown, the gate reduction increases with increasing time-step for all system sizes, which implies that the circuit-reduction benefit of the DS compiler scales well with simulation time-step.  Furthermore, for each time-step, the number of reduced gates increases with additional qubits, indicating that the circuit-reduction benefit of the DS compiler also scales well with growing system size.  Figure \ref{fig:rigetti_performance}b shows the number of reduced gates as a percentage of the circuit size produced by Rigetti's general-purpose compiler.  The percent reduction in native gate count asymptotes with growing time-step for all system sizes in a range around 30\%.  Figure \ref{fig:rigetti_performance}c shows the percent reduction in wall-clock compilation time of the DS compiler compared to Rigetti's general-purpose compiler, which asymptotes to values ranging within 30-60\%.    

\subsection{IBM Native Gate Set Domain-Specific Compiler}
\begin{algorithm}[!h]
\caption{Pseudocode for domain-specific compilation of TFIM circuits into IBM's native gate set.}
 Read in high-level gates\;
 {\scriptsize
 1: \While{
 $\Qcircuit @C=1em @R=.7em {
    & \ctrl{1} & \qw\\ 
    &\targ     &  \qw
    }$ exists } { \vspace{0.5cm}
    replace it with
   $\Qcircuit @C=1em @R=.7em {
    & \gate{H} &  \targ & \gate{H} & \qw\\ 
    & \gate{H} & \ctrl{-1}  &  \gate{H} & \qw }$
    } 
    
 2: \While{
 $\Qcircuit @C=1em @R=.7em {
    & \gate{H}& \gate{H} & \qw
    }$ exists }{\vspace{0.5cm}
    replace it with
  $\Qcircuit @C=1em @R=.7em {
    & \qw & \qw & \qw & \qw} $ 
    }  
    
 3: \While{
 $\Qcircuit @C=1em @R=.7em {
    & \gate{H}& \gate{RZ(\theta)} &\gate{H} & \qw
    }$ exists }{\vspace{0.5cm}
    replace it with
  $\Qcircuit @C=1em @R=.7em {
    & \gate{RZ(\frac{\pi}{2})} & \gate{RX(\frac{\pi}{2})} 
    & \gate{RZ(\theta + \pi)} & \gate{RX(\frac{\pi}{2})}
    & \gate{RZ(\frac{\pi}{2})} & \qw} $ 
    
    4: \While{
    $\Qcircuit @C=1em @R=.7em {
    & \qw & \targ & \qw \\
    & \gate{RZ(\theta_i)} & \ctrl{-1} & \gate{RZ(\theta_j)}
    }$ exists }{\vspace{0.5cm}
    replace it with
  $\Qcircuit @C=1em @R=.7em {
    & \targ & \qw \\
    & \ctrl{-1} & \gate{RZ(\theta_i + \theta_j)}} $ 
    }
   }
 5: \While{
 $\Qcircuit @C=1em @R=.7em {
    & \gate{H}& \gate{RZ(\theta)} & \qw
    }$ exists }{\vspace{0.5cm}
    replace it with
  $\Qcircuit @C=1em @R=.7em {
    & \gate{RZ(\frac{\pi}{2})} & \gate{RX(\frac{\pi}{2})}
    & \gate{RZ(\theta + \frac{\pi}{2})} & \qw} $ 
 
    6: \While{$    \Qcircuit @C=1em @R=.7em {
    & \gate{RZ(\theta_{i})} & \gate{RZ(\theta_{j})} & \qw
    } $ exists }{\vspace{0.5cm}
    replace it with
      $ \Qcircuit @C=1em @R=.7em {
    & \gate{RZ(\theta_{i}+\theta_{j})} & \qw }$
     }
    }
}
 \label{ibm_pseudocode}
\end{algorithm}
 
Algorithm \ref{ibm_pseudocode} shows pseudocode detailing the order in which a subset of the identities from Table \ref{gate_identites} are applied in the DS compiler for IBM's native gate set.  The first loop applies identity 3 to swap the control and target qubits of each $CNOT$ gate, which primes the circuit for further reductions.  The second loop uses identity 6 to remove consecutive pairs of Hadamard gates acting on the same qubit.  The third loop searches for an $RZ(\theta)$ gate sandwiched between two Hadamard gates, a common motif found in the input circuits, and applies identities 3, 8, and 10 in concert with its inner loop (fourth loop) to transform all Hadamard gates into native gates, contract consecutive rotation gates about the same axis on the same qubit, and contract $RZ$ gates across the control qubits of $CNOT$ gates.  The fifth loop transforms any remaining Hadamard gates to native gates, working in concert with its inner loop (sixth loop) to contract pairs of consecutive $RZ$ gates on the same qubit present in the circuit.  Finally, not shown in the pseudocode, is the removal of any trailing gates at the end of the circuit that only alter the phase of the system, and thus do not affect a measurement gate that immediately follows.  Like in Algorithm 1, a heuristic similar to the technique of unit propagation is applied in Algorithm 2 in the third and fifth loops, both of which are outer loops of gate rearrangement with inner loops of gate reduction.
\begin{figure}[!h]
\centering
\includegraphics[scale=0.85]{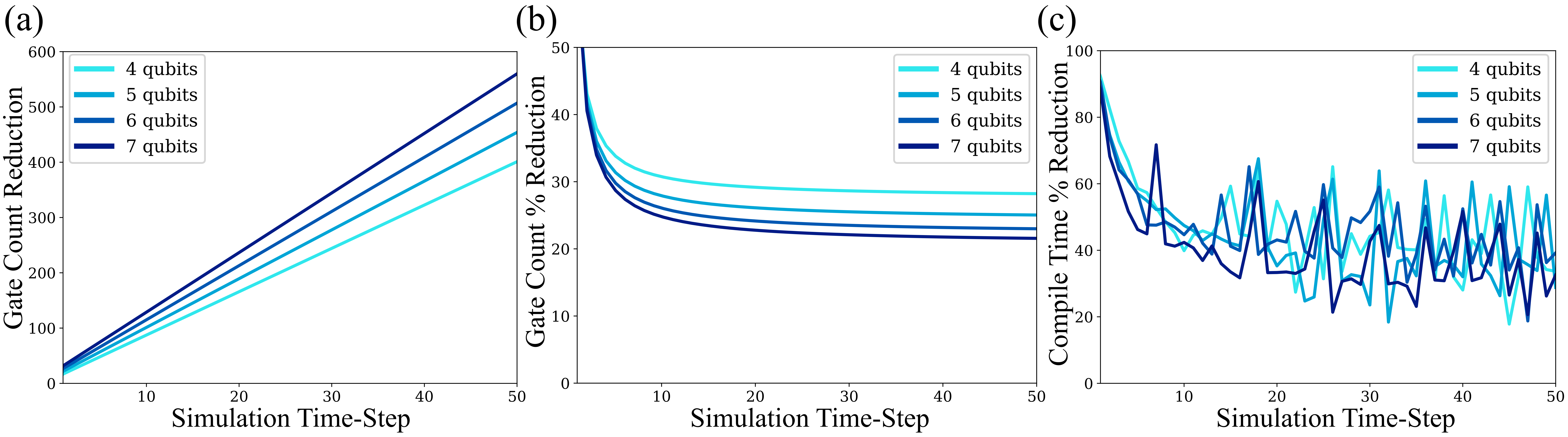}
\caption{Performance comparison of the DS compiler to IBM's general-purpose compiler. (a) Absolute reduction in gate count using the DS compiler over IBM's general-purpose compiler, for varying system size.  (b) Percent reduction in gate count using the DS compiler with respect to IBM-compiled circuit size, for varying system size.  (c) Percent reduction in wall-clock compilation time using the DS compiler with respect to IBM-compiled circuit size, for varying system size.}
\label{fig:ibm_performance}
\end{figure}
To test the performance of this compiler, we used the same simulation algorithm used for testing the Rigetti compilers to create a sequence of high-level TFIM circuits.  Results can be found in Figure \ref{fig:ibm_performance}.  Figure \ref{fig:ibm_performance}a shows the absolute native gate count reduction when using the DS compiler over IBM's general-purpose compiler.  Again, the gate reductions increase with increasing time-step for all system sizes, indicating that the circuit-reduction benefit of the DS compiler scales well with simulation time-step; furthermore, for each time-step, the number of reduced gates increases with additional qubits, indicating that the circuit-reduction benefit of the DS compiler also scales well with growing system size.  Figure \ref{fig:ibm_performance}b shows the number of reduced gates as a percentage of the circuit size produced by IBM's general-purpose compiler.  The percent reduction in gate count asymptotes with growing time-step for all system sizes in a range around 25\%.  Figure \ref{fig:ibm_performance}c shows the percent reduction in wall-clock compilation time of the DS compiler compared to IBM's general-purpose compiler, which asymptotes to values ranging within 20-60\%.
 
\section{Discussion}
While sufficient quantum hardware exists for demonstrating quantum supremacy with random circuits\cite{arute2019quantum}, a major bottleneck to demonstrating physical quantum supremacy is the development of new quantum circuit compilers for optimizing such circuits.  Though numerous software packages with general-purpose compilers have recently been developed\cite{wecker2015liqui, smith2016practical, paler2017fault, haner2018software}, the development of DS compilers, optimized for certain classes of circuits and for specific native gate sets, may be a necessary intermediate technology for the success of early NISQ-era simulations.  The two DS compilers we developed for currently available NISQ computers, using heuristics derived from artificial intelligence (AI) techniques, demonstrated 25-30\% reductions in circuit size compared to the general-purpose compilers.   This significant reduction in circuit size may soon allow relevant dynamic simulations of quantum materials on near-future NISQ computers, providing new insights from results not achievable with classical computers.  In future work, we would like to explore the use of other optimization tools developed in AI. Several tools for solving SAT, constraint satisfaction, and weighted constraint satisfaction problems are currently being used to solve many real-world problems after proper reformulation. These tools encapsulate many decades of AI research for optimization and work well in practice for NP-hard problems. Some attempts have already been made for minimizing the number of gates in digital logic circuits by invoking powerful SAT solvers~\cite{stc03}.  In addition TFIM circuits, DS compilers can also be created for other classes of quantum circuits for dynamic simulation\cite{martinez2016real, zhukov2018algorithmic, smith2019simulating}.  

\section{Methods}
\subsection{Decomposition of Time-Evolution Operator}
In order to map the time-ordered exponential unitary operator into a set of one- and two-qubit gates, two approximations must be applied\cite{poulin2011quantum}.  First, the time-dependence of $H(t)$ must be ignored on time scales smaller than some chosen, minimal time-step $\Delta t$.  The Hamiltonian can then be approximated as a piece-wise constant function that takes the constant value $H((j+\frac{1}{2})\Delta t)$ during the time interval $[j\Delta t, (j+1)\Delta t]$, where $j$ is some integer, resulting in the following decomposition: $U(n\Delta t) \approx \prod_{j=0}^{n-1} e^{-iH((j+\frac{1}{2})\Delta t)\Delta t}$.  Second, each matrix exponential in this product must be approximated with the Trotter decomposition\cite{trotter1959product}.  To perform the Trotter decomposition, the Hamiltonian is divided into components, each of which is efficiently diagonalizable.  For the TFIM Hamiltonian, this can be accomplished with the following decomposition: $H(t)= H_x(t) + H_z$ where $H_x(t)=-B(t)\sum_{i=1}^{N} \sigma_{i}^{x}$ and $H_z = -J_z \sum_{i=1}^{N-1} \sigma_{i}^{z} \sigma_{i+1}^{z}$. Thus, the time evolution operator is finally approximated as: $U(n\Delta t) = \Sigma_{j=0}^{n-1} e^{-iH_{x}((j+1/2)\Delta t)\Delta t} e^{iH_{z}\Delta t} + \mathcal{O}(\Delta t)$, as was given in Equation \ref{unitary}.

\subsection{Algorithm for Circuit Generation}
An algorithm was written to generate circuits that simulate the time evolution of a system of spins under the TFIM Hamiltonian given in Equation \ref{hamiltonian}.  For a simulation to time $n\Delta t$, $n$ different circuits are created.  A physically reasonable value for material systems was chosen for $J_z$, and a sinusoidal function was chosen for $B(t)$, with amplitude and frequency also assigned physically reasonable values.  For a given circuit, the algorithm proceeds by appending alternating sets of gates that each propagate the system forward by a time-step $\Delta t$ according to the two exponentials given in the product in Equation \ref{unitary}.  Note that the parameters in the set of gates carrying out the exponential $e^{-iH_x((j+\frac{1}{2})\Delta t)\Delta t}$ will change for each subsequent propagation by $\Delta t$ because our Hamiltonian in Equation \ref{hamiltonian} is time-dependent.  The set of gates representing the exponential  $e^{-iH_z\Delta t}$ is identical for each application.  Once gate sets representing the two exponentials have been alternately applied $n$ times each, the circuit simulating to time $n\Delta t$ is complete and the algorithm begins building the circuit for simulation to the subsequent time-step.  Circuits for all time-steps are appended to a list, which serves as input to the various compilers for conversion to native gates and compression.

\subsection{Development and Performance Analysis of Compilers}
Written in the Python programming language, both compilers were developed to reduce circuit size for circuits designed to simulate dynamic evolution of TFIM systems.  The compilers take as input a list of circuits comprised of high-level gates represented as a serial list of gates and the qubits upon which they act.  The compilers output a list of circuits of reduced size comprised solely of native gates.  Since the native gate sets for Rigetti and IBM differ, separate compilers were developed for each platform.  

Two points should be mentioned about our compilers.  First, note that the compilers take advantage of the fact that for these dynamic simulations, only measurement probabilities of different states need to be conserved when applying transformation and compression identities.  This means that the overall global phase of the system, which has no observable effect on measurement probabilities, need not be conserved, allowing for a larger number of useful circuit identities to be used.  Second, we note that the compilers are deterministic in their approach; a given high-level circuit will be compiled to the same executable circuit on each run.  This is as opposed to other available compilers which return different compiled circuits, of varying size, for the same high-level circuit on different runs.

Both Rigetti's general-purpose compiler and our DS compiler compile circuits into Rigetti's native gate set.  To compare performance of the two compilers in terms of circuit size, we therefore counted the number of native gates in the circuits produced for each time-step by each compiler.  One complication is that Rigetti's compiler is not deterministic, meaning that running the same input circuit through the compiler two different times can result in two different output circuits.  To remove such fluctuations, we ran Rigetti's general-purpose compiler three times and averaged the number of native gates in the circuits over the three runs.  These averaged native gate counts for each time step were used for comparison with the native gate counts from our deterministic DS compiler.  For comparing compiler run times, wall-clock times were recorded for three separate runs each of Rigetti's compiler and our DS compiler, and the average over these runs was used for comparison.

Performance comparison in terms of native gate count was slightly more subtle for the IBM compilers.  IBM's general-purpose compiler compiles circuits into a set of basis gates, comprised of {$U1(\lambda), U2(\phi, \lambda), U3(\theta, \phi, \lambda), CNOT$}.  The $U1, U2, U3$ basis gates can be written in terms of native gates as shown in Table \ref{ugates}.  Even though its general-purpose compiler compiles to this basis gate set, the IBM quantum computer nonetheless only physically executes gates from the native gate set.  For this reason, we chose to implement our DS compiler to output circuits comprised of native gates.
\begin{table}
\caption{IBM Basis and Native Gate Equivalents}
\label{pulse_identities}
\begin{tabular}{|c|c|}
\hline \hline
Basis Gate & Native Gate Equivalent\\ \hline
$\Qcircuit @C=1em @R=.7em {
    & \gate{U1(\lambda)} & \qw
    }$ & $\Qcircuit @C=1em @R=.7em {
    & \gate{RZ(\lambda)} & \qw
    }$ \\ \hline
$\Qcircuit @C=1em @R=.7em {
    & \gate{U2(\phi,\lambda)} & \qw
    }$ & $\Qcircuit @C=1em @R=.7em {
    & \gate{RZ(\phi + \frac{\pi}{2})} & \gate{RX(\frac{\pi}{2})} & \gate{RZ(\lambda - \frac{\pi}{2})} & \qw } $ \\ \hline
$\Qcircuit @C=1em @R=.7em {
    & \gate{U3(\theta,\phi,\lambda)} & \qw
    }$ & $\Qcircuit @C=1em @R=.7em {
    & \gate{RZ(\phi + 3\pi)} & \gate{RX(\frac{\pi}{2})} & \gate{RZ(\theta + \pi)} & \gate{RX(\frac{\pi}{2})} & \gate{RZ(\lambda)} & \qw } $ \\ \hline
\hline \hline
\end{tabular}
\label{ugates}
\end{table}
As shown in Table \ref{ugates}, different basis gates are comprised of different numbers of native gates.  Therefore, to fairly compare the circuit sizes produced by IBM's general-purpose compiler to those output by our DS compiler, we counted all $U1$ and $CNOT$ gates in each IBM-compiled circuit as one native gate each, $U2$ gates as three native gates, and $U3$ gates as five native gates.  All gate count comparisons for the IBM compilers, therefore, were in terms of native gate count.  This also makes the gate count performance of the IBM compilers easier to compare with the gate count performance of the Rigetti compilers.

\section*{Acknowledgements}
This work was supported as part of the Computational Materials Sciences Program funded by the U.S. Department of Energy, Office of Science, Basic Energy Sciences, under Award Number DE-SC0014607.

\section*{Code Availability}
The two domain-specific compilers are available in the online repository \url{https://github.com/sgulania/DSQC}.  The code to create the TFIM circuits used to test the compilers are available in the online repository \url{https://github.com/lebassman/TFIM_Trotter_Simulations}.

\newpage
\bibliographystyle{naturemag}
\bibliography{qc}
\end{document}